\newif\ifCHANGES
\let\ifCHANGES\iftrue

\documentclass[
    5p,
    sort&compress,
    times,
]{elsarticle}

\usepackage[utf8]{inputenc}

\usepackage{xparse}

\usepackage{microtype}

\usepackage{amsmath}
\usepackage{amssymb}
\usepackage{mathtools}
\usepackage{bm}
\usepackage{physics}
\usepackage{siunitx}

\usepackage{graphicx}
\usepackage{xcolor}

\usepackage{todonotes}

\usepackage{acronym}
\usepackage{hyperref}

\definecolor{edits}{RGB}{220,0,0}
\definecolor{strike}{RGB}{150,50,50}
\ifCHANGES
    \usepackage[normalem]{ulem}
    
    \NewDocumentCommand\STRIKE{+m}{{\color{strike}\sout{#1}}}
\else
    
    \NewDocumentCommand\STRIKE{+m}{}
\fi

\NewDocumentCommand\ie{}{i.\,e.}
\NewDocumentCommand\cf{}{cf.}

\NewDocumentCommand\mathperiod{}{\,\text{.}}
\NewDocumentCommand\mathcomma{}{\,\text{,}}

\RenewDocumentCommand\vec{m}{\bm{#1}}
\NewDocumentCommand\mat{m}{\bm{#1}}

\NewDocumentCommand\kB{}{k_\mathrm{B}}

\NewDocumentCommand\traj{}{\ensuremath{\gamma}}
\NewDocumentCommand\trajNhim{}{\ensuremath{\traj^\ddagger}}
\NewDocumentCommand\kFloq{}{\ensuremath{\bar{k}_\mathrm{F}}}
\NewDocumentCommand\reacEns{}{\ensuremath{\chi_\mathrm{r}^\mathrm{e}}}
\NewDocumentCommand\EEnsMin{}{\ensuremath{E_\mathrm{th}^\mathrm{e}}}

\NewDocumentCommand\ENhim{}{\ensuremath{E_{\trajNhim}}}
\NewDocumentCommand\ENhimAvrg{s}
    {\ensuremath{\IfBooleanTF{#1}{\ev*}{\ev}{\ENhim}_t}}
\NewDocumentCommand\ESrc{}{\ensuremath{E_{-}^\mathrm{b}}}
\NewDocumentCommand\ESpMin{}{\ensuremath{E_\mathrm{th}^\mathrm{b}}}
\NewDocumentCommand\ESpMinTmAvrg{s}
    {\ensuremath{\IfBooleanTF{#1}{\ev*}{\ev}{\ESpMin}_{t_0}}}
\NewDocumentCommand\ESpMinTmMin{}{\ensuremath{\min_{t_0} \ESpMin}}
\NewDocumentCommand\survival{mm}{\ensuremath{\mathrm{SF}_{#1}^{(#2)}}}

\journal{Physica D}

\begin{document}

\begin{frontmatter}
    \title{Controlling reaction dynamics in chemical model systems
        through external driving}

    \author[us]{Johannes Reiff}
    \author[us]{Robin Bardakcioglu}
    \author[us]{Matthias Feldmaier}
    \author[us]{Jörg Main}
    \address[us]{
        Institut für Theoretische Physik I,
        Universität Stuttgart,
        70550 Stuttgart, Germany
    }

    \author[jhuc,jhucbe]{Rigoberto Hernandez\texorpdfstring{\corref{cor}}{}}
    \ead{r.hernandez@jhu.edu}
    \cortext[cor]{Corresponding author}
    \address[jhuc]{
        Department of Chemistry,
        Johns Hopkins University,
        Baltimore, Maryland 21218, USA
    }
    \address[jhucbe]{
        Departments of Chemical \& Biomolecular Engineering,
        and Materials Science and Engineering,
        Johns Hopkins University, \\
        Baltimore, Maryland 21218, USA
    }

    \date{\today}

    \begin{abstract}
        The rate of a chemical reaction can often be determined by
        the properties of a rank-1 saddle
        and the associated transition state separating reactants and products.
        We have found evidence that such rates
        can be controlled and even enhanced by external driving
        in at least one such system.
        Specifically, we analyze a reactive model in two degrees of freedom
        that has been used earlier to describe driven chemical reactions.
        Therein, changes in the external driving
        can lead to a local maximum of the decay rate constant
        or even to bifurcations of periodic trajectories on
        the normally hyperbolic invariant manifold (NHIM)
        corresponding to the transition state.
        Inspired by these bifurcations, we show that
        in this case, the dynamics on the NHIM
        can be connected to the geometry of reactive trajectories
        and to reaction probabilities
        of Maxwell--Boltzmann distributed reactant ensembles.
    \end{abstract}

\end{frontmatter}


\acrodef{DS}{dividing surface}
\acrodef{NHIM}{normally hyperbolic invariant manifold}
\acrodef{PSOS}{Poincaré surface of section}
\acrodef{TS}{transition state}
\acrodef{TST}{transition state theory}


\section{Introduction}
\label{sec:intro}

Reactant and product states in a chemical reaction are usually separated
by a barrier that needs to be surmounted during the reaction.
Dynamics near the barrier have been successfully described
by the framework of \ac{TST}~\cite{
    eyring35, wigner37, pech81, truh96, dawn05a, dawn05b, peters14a, wiggins16}.
Beyond the reactants and products,
\ac{TST} focuses on the determination of the \ac{TS},
an unstable state confined indefinitely at the transition barrier
that is neither reactant nor product.
Alternatively, one could focus on the correlation
between the respective fluxes through the dividing surfaces
associated with the reactants and products
as they enter the flux correlation formalism for the rate formula~\cite{mill98}.
The determination of the transition paths between such surfaces
has recently been used as the basis of accurate rate formulas
in a non-driven system~\cite{makarov18}.
Here, instead, we focus on the use of the \ac{TS}
as the barrier separating reactants from products.
In arbitrary dimensions, the \ac{TS} is embedded in
the \ac{NHIM} of the associated barrier region
separating reactants from products~\cite{hern19e, Naik2019b}.
As the name suggests,
reactants need to pass close to this \ac{TS} in order to react.
Therefore, it should not be surprising that
a \ac{DS} separating reactants from products
can be attached to the \ac{NHIM}%
~\cite{wiggins01, Uzer02, dawn05a, dawn05b, Ezra2018a}.

Recent advances in the use of \acp{TS} to obtain chemical reaction rates
include the resolution of classical model systems~\cite{
    dawn05a, komatsuzaki06a, Komatsuzaki2010, komatsuzaki11, waalkens13,
    Maug2013, komatsuzaki15b, wiggins16, lorquet17, waalkens18, komatsuzaki18,
    keshavamurthy18, Naik2019a}
of varying complexity
as well as quantum mechanical problems~\cite{pollak17}.
Most of these problems feature transitions over a rank-1 saddle
along a confined reaction pathway in a time-independent system
invariably using perturbative expansions.
As not all systems admit to such solutions,
we have also pursued alternate approaches,
including Lagrangian descriptors~\cite{hern15e, hern16a, hern17h},
binary contraction~\cite{hern18g},
and machine learning~\cite{hern18c}.
In this paper,
we address the emergent dynamics of a time-dependent chemical model system
under periodic external driving of the transition barrier.
We find in Secs.~\ref{sec:results/nhim} and~\ref{sec:results/rate}
that our model admits to a decay rate
that is highly sensitive to the strength and frequency of the driving,
and hence the decay rate can be controlled.
In the process,
the structure of the \ac{NHIM} changes qualitatively via bifurcations.
Section~\ref{sec:results/geometry} further analyzes one such bifurcation
and how it influences reactive trajectories passing close to the \ac{NHIM}.
The results allow us to predict reaction probabilities
in Sec.~\ref{sec:results/probability}.
These sections clarify
how the dynamics on the \ac{NHIM} translates
to reactive properties of ensembles starting far from it.


\section{Materials and methods}
\label{sec:methods}

We represent \textit{driven} chemical reactions using a model explored
in previous work~\cite{hern17h, hern18c, hern18g, hern19a, hern20d, hern20n}.
As in typical chemical reactions,
the barrier region is represented as a rank-1 saddle
separating reaction and product basins
such as shown in the contour plots of Fig.~\ref{fig:potential}.
The driving arises by way of coupling between a time-dependent external field
and the dipole associated with the reaction coordinate~\cite{borondo10, hern15a}.
Besides being physically relevant to chemical reactions,
the restriction to unbounded reactant and product basins is a simplification that
avoids the global recrossings that would arise if one or both basins were closed.
Nevertheless,
we emphasize that the presence of closed reactant and product basins
would not challenge the methods presented here
because the important dynamics is happening in the saddle region.
For more information on how to deal with global recrossings
see Ref.~\cite{hern17e}.


\subsection{Model potential}
\label{sec:methods/model}

\begin{figure}
   \includegraphics[width=\columnwidth]{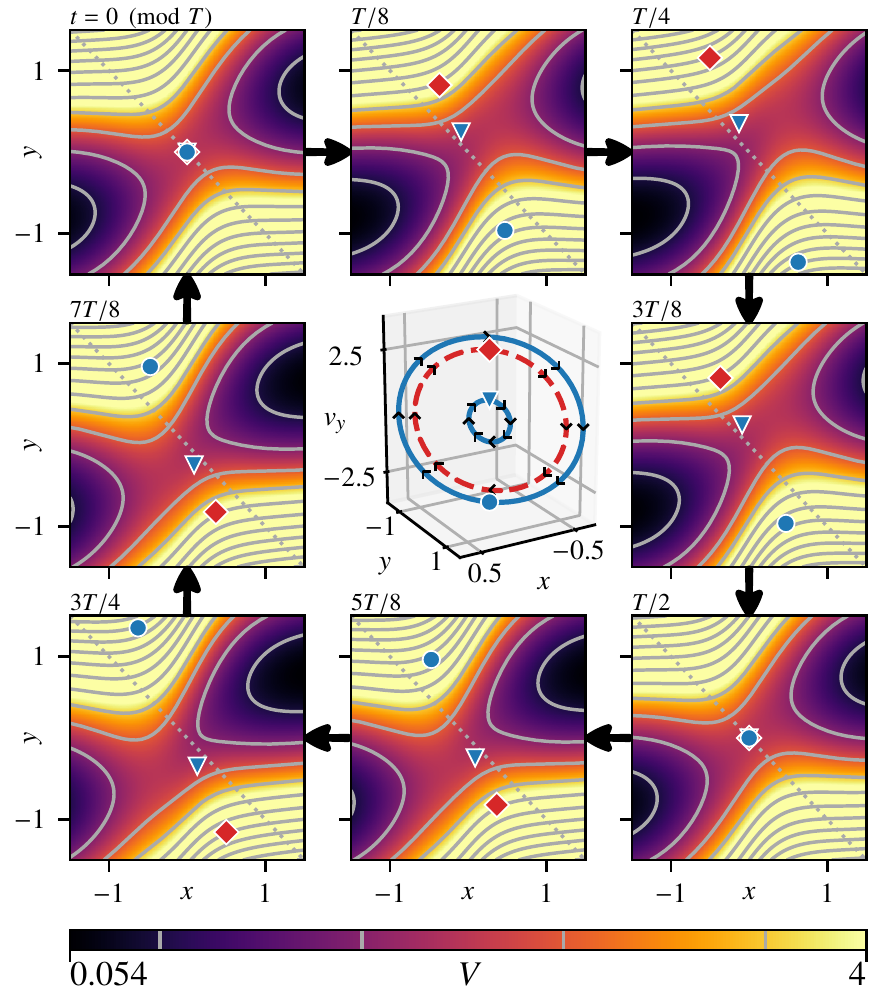}
    \caption{%
        Outer ring:
        Snapshots of the time-dependent potential $V(x, y, t)$
        following Eq.~\eqref{eq:potential}
        with $\omega_x = \num{0.77}\,\pi$ and $\hat{x} = \num{0.4}$
        over the oscillation period $T = 2 \pi / \omega_x$.
        Diamond, triangle, and circle markers symbolize
        the position of three particles with trajectories of period $T$.
        Equidistant contour lines with $\Delta V = 1$
        are shown in solid gray.
        A diagonal dotted line per panel serves as a guide to the eye,
        elucidating the saddle movement.
        The color scale is capped at $V = 4$.
        Center:
        Aforementioned periodic trajectories in phase space.
        Filled markers symbolize positions at $t = 0$
        while positions of the particles at the potential snapshots
        are indicated by black arrow heads pointing in the direction of motion.}
    \label{fig:potential}
\end{figure}

In more detail, we use a two-dimensional potential
\begin{equation}
    \begin{split}
        V(x, y, t)
            &= E_\mathrm{b} \exp(-\qty[x - \hat{x} \sin(\omega_x t)]^2) \\
            &\quad + \frac{\omega_y^2}{2} \qty[y - \frac{2}{\pi} \arctan(2 x)]^2
        \mathcomma
    \end{split}
    \label{eq:potential}
\end{equation}
where a time-periodically oscillating Gaussian barrier with height $E_\mathrm{b}$
separates an unbounded reactant from an unbounded product basin,
as shown in Fig.~\ref{fig:potential}.
The barrier moves along the $x$-coordinate
with frequency $\omega_x$ and amplitude $\hat{x}$.
In the $y$-direction, the particles are bound
by a harmonic potential with frequency $\omega_y$.
To make the system nonlinear, the $x$ and $y$ coordinates are coupled
so that the minimum energy path for a reaction over the saddle is given by
$y = (2 / \pi) \arctan(2 x)$.
For simplicity, we use dimensionless units with parameters
$E_\mathrm{b} = \num{2}$ and $\omega_y = \num{2}$.
The aim of this work is to demonstrate
how chemical reactions can be controlled by external driving.
We focus on the dependence of the system
on the parameters $\omega_x$ and $\hat{x}$
since they describe the saddle movement caused by the external driving.


\subsection{Decay rates}
\label{sec:methods/rates}

To calculate the decay rate associated with trajectories on the \ac{NHIM},
we use the Floquet method
first introduced in Ref.~\cite{hern14f}
and later extended in Refs.~\cite{hern17f, hern19e, hern20d}.
The method relies on the conjecture%
---verified under certain assumptions such as those shown here---%
that the decay rate of reactants into products near a \ac{TS}
is related to the Floquet coefficients of the \ac{TS}.
It exploits the fact that trajectories near a periodic orbit on the \ac{NHIM}
can be described using a linearization of the equations of motion.
Let $\Delta \vec{\traj}(t)$ represent the deviation of a trajectory
from the orbit with period $T_\mathrm{po}$.
Its time evolution can then be described by
\begin{equation}
    \label{eq:linear_eom}
    \Delta \dot{\vec{\traj}}(t) = \mat{J}(t) \Delta \vec{\traj}(t)
\end{equation}
where $\mat{J}(t)$ is the system's Jacobian evaluated on the periodic orbit.
By leveraging its linearity, Eq.~\eqref{eq:linear_eom} can also be expressed as
\begin{equation}
    \Delta \vec{\traj}(t) = \mat{\sigma}(t) \Delta \vec{\traj}(0)
\end{equation}
where the fundamental matrix $\mat{\sigma}(t)$ is defined by
\begin{equation}
    \dot{\mat{\sigma}}(t) = \mat{J}(t) \mat{\sigma}(t)
    \qq{with}
    \mat{\sigma}(0) = \mat{1}
\end{equation}
and $\mat{1}$ being the identity matrix.
The rate constant \kFloq\ then follows from
the largest and smallest eigenvalue $m_\mathrm{u, s}(t)$ of $\mat{\sigma}(t)$ as
\begin{equation}
    \label{eq:floquet_rate}
    \kFloq T_\mathrm{po}
        = \ln|m_\mathrm{u}(T_\mathrm{po})| - \ln|m_\mathrm{s}(T_\mathrm{po})|
    \mathperiod
\end{equation}
This method can be generalized to non-periodic trajectories
by evaluating the right-hand side of Eq.~\eqref{eq:floquet_rate}
for sufficiently long times and then
applying a linear regression~\cite{hern19e}.


\section{Results and discussion}
\label{sec:results}

We start by investigating two examples in which the external driving
can be seen to affect the dynamics of the activated complex.
In the following, we use $\ev{X}_Y$ to denote
the average of quantity $X(Y)$ over $Y$.


\subsection{Dynamics on the NHIM}
\label{sec:results/nhim}

\begin{figure}[t]
    \includegraphics[width=\columnwidth]{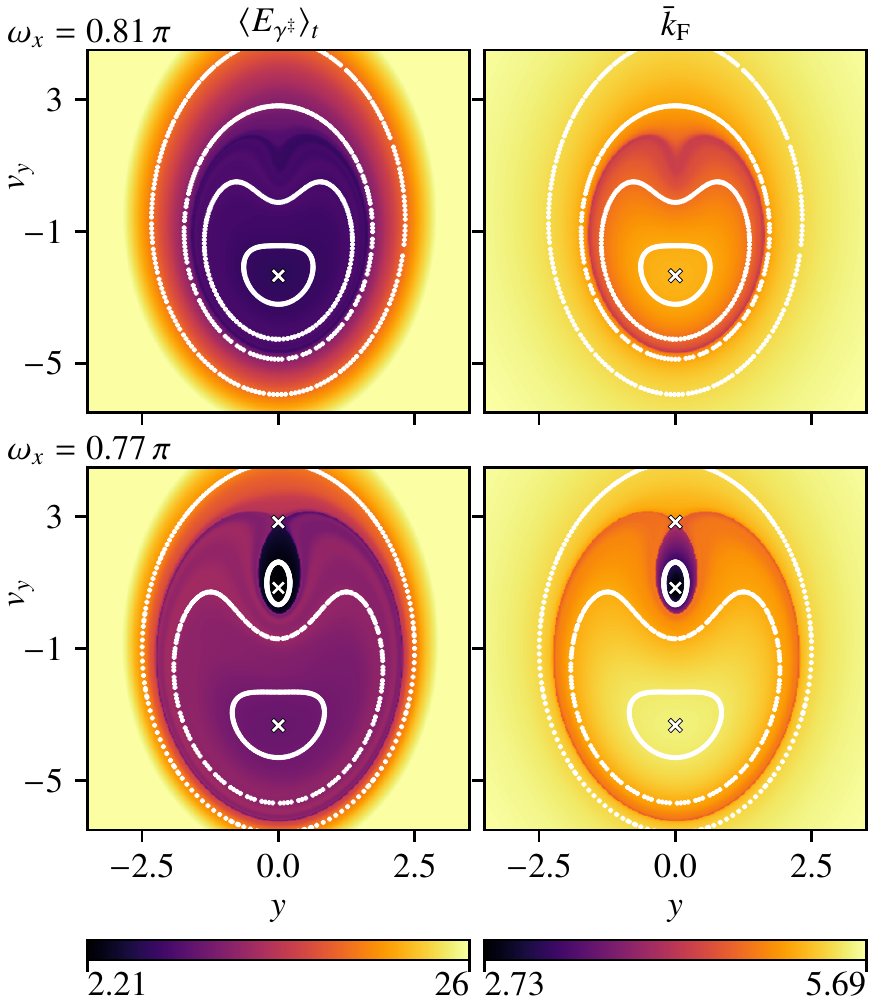}
    \caption{%
        Average total energy \ENhimAvrg*
        (left column, capped at $\ENhimAvrg* = \num{26}$)
        and rate constant \kFloq\ (right column)
        of trajectories \trajNhim\
        started at position $(y, v_y)$ and time $t_0 = 0$
        on the \acs{NHIM}.
        Shown are typical plots at $\hat{x} = \num{0.4}$
        before (top row, $\omega_x = \num{0.81}\,\pi$)
        and after (bottom row, $\omega_x = \num{0.77}\,\pi$) the bifurcation.
        To elucidate the structure, a stroboscopic map of selected trajectories
        (\acs{PSOS}, white dots) has been overlaid
        highlighting the existence of one elliptic fixed point before
        and two elliptic as well as one hyperbolic fixed point after
        the bifurcation (small white crosses).
        While the elliptic fixed points are
        always characterized by local minima in \ENhimAvrg*,
        they feature either a local minimum or maximum in \kFloq.}
    \label{fig:energy_rate_psos}
\end{figure}

The left column of Fig.~\ref{fig:energy_rate_psos}
shows the time-averaged total energy \ENhimAvrg*
of trajectories \trajNhim\ on the \ac{NHIM}
for two values of $\omega_x$.
At $\omega_x = \num{0.81}\,\pi$,
\ENhimAvrg* reveals a region in phase space with low-energy trajectories.
This region is accompanied by a local maximum in the decay rate \kFloq,
as can be seen in the right column of Fig.~\ref{fig:energy_rate_psos}.
The overlaid \ac{PSOS} highlights an elliptic fixed point
belonging to the associated periodic orbit.
This orbit fulfills all requirements for a \ac{TS} trajectory
as defined in Refs.~\cite{hern19a, hern19e, hern20d}.
It can be seen as the dominant trajectory in the sense
that decay rates from this trajectory are also characteristic
of neighboring trajectories.

For decreasing $\omega_x$, two new fixed points
and, hence, periodic trajectories emerge,
as can be seen in Fig.~\ref{fig:energy_rate_psos} at $\omega_x = \num{0.77}\,\pi$.
These trajectories are shown in Fig.~\ref{fig:potential}
for the elliptic (solid blue lines)
and hyperbolic (dashed red line) fixed points.
This so-called
saddle-node bifurcation~\cite{borondo95a, borondo96a, Li09, Inarrea2011, hern20n}
qualitatively changes the dynamics on the \ac{NHIM}.
While the new elliptic fixed point
is still characterized by a local minimum in \ENhimAvrg*,
it now also features a local minimum in \kFloq\ instead of a maximum.
In addition, its energy \ENhimAvrg* is much lower
compared to the original elliptic fixed point at
velocity $v_y = \dot{y} \approx \num{-3}$.

\begin{figure}[t]
    \includegraphics[width=\columnwidth]{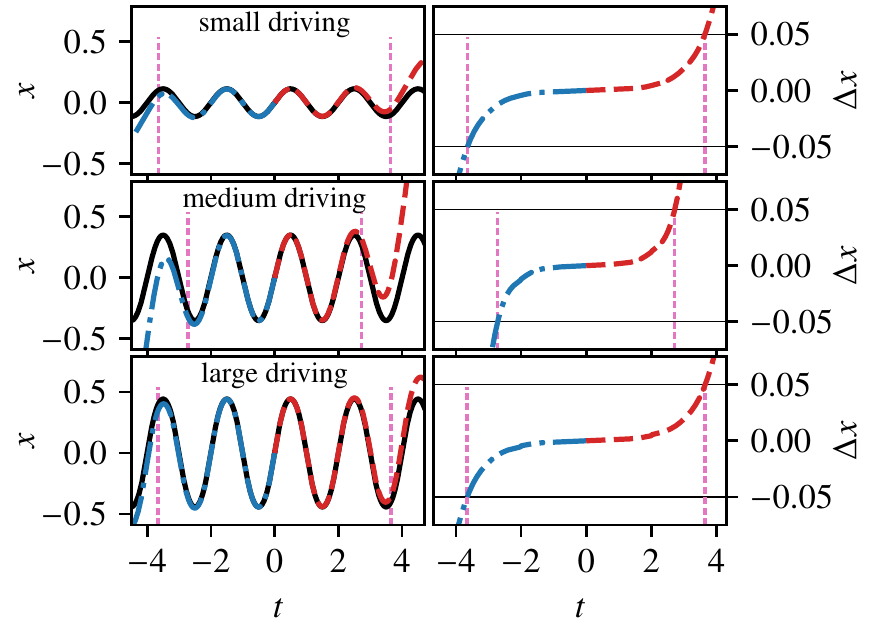}
    \caption{%
        Position $x$ over time $t$ for three systems
        with small ($\hat{x} = \num{0.1}$),
        medium ($\hat{x} = \num{0.6}$),
        and large ($\hat{x} = \num{1.6}$) driving.
        Driving frequency in all cases is $\omega_x = \pi$.
        For each parameter set, the periodic trajectory (black solid)
        as well as a trajectory offset by $\Delta v_x = \num{e-3}$
        at $t_0 = 0$ (blue dash dotted/red dashed) are shown.
        The stability of the trajectories is indicated in the right column
        by the time interval (vertical lines)
        where the deviation $\Delta x$ from the \acs{TS}
        is less than \num{0.05} (horizontal lines).
        Stability is minimal for medium driving.}
    \label{fig:trajectories}
\end{figure}

The second example for
the influence of external driving on the dynamics on the \ac{NHIM}
is illustrated in Fig.~\ref{fig:trajectories}.
At a fixed $\omega_x = \pi$,
only a single periodic trajectory was found in the examined parameter regime.
Particles near this periodic trajectory
exhibit a change in their stability
that depends on the system's driving amplitude $\hat{x}$.
We illustrate the change in the
stability using the time it takes the particle to reach
a distance of $\abs{\Delta x} = \num{0.05}$ from the periodic trajectory.
When only a small driving is applied,
the particle stays in the saddle region for a relatively long time.
This in turn indicates a low decay rate \kFloq.
When increasing the amplitude, stability initially decreases for medium driving
only to increase again for large driving.
As a result, there must be a local maximum in the systems decay rate \kFloq\
allowing for rate enhancement through optimization of $\hat{x}$.


\subsection{Decay rate enhancement}
\label{sec:results/rate}

\begin{figure}[t]
    \includegraphics[width=\columnwidth]{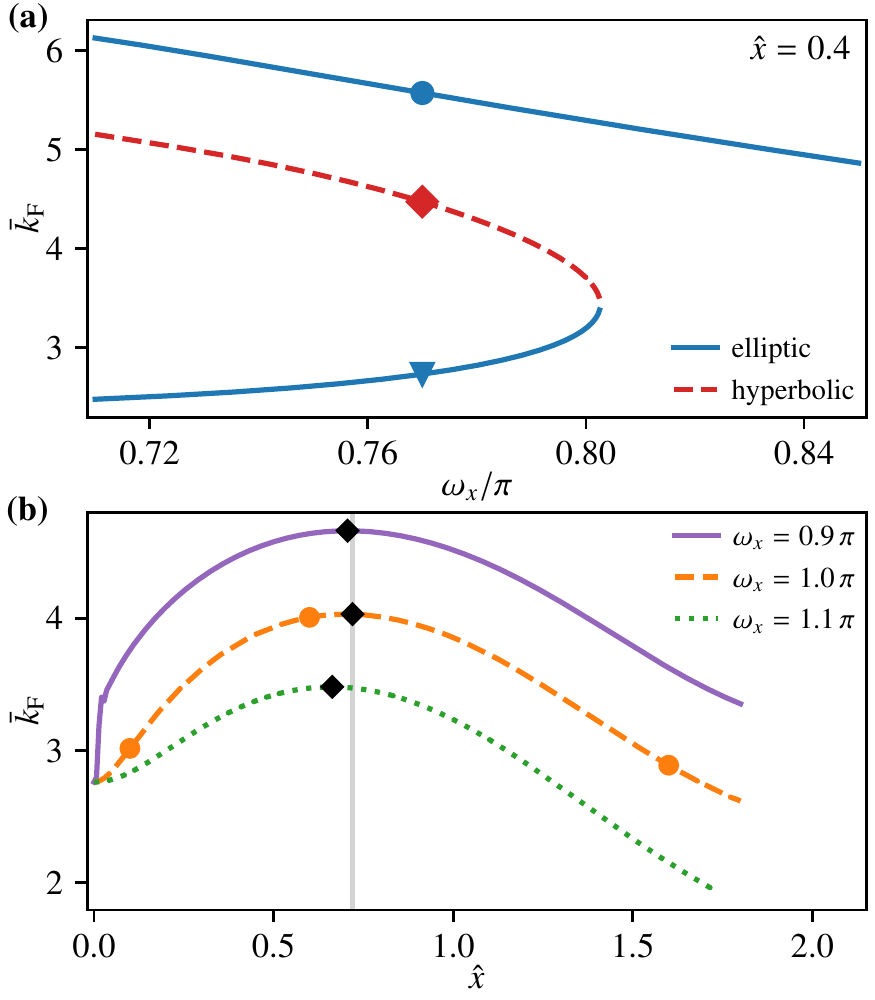}
    \caption{%
        (a)~Rate constant \kFloq\ as a function of driving frequency $\omega_x$
        for two elliptic and one hyperbolic fixed point,
        each corresponding to a periodic trajectory.
        At $\omega_x \approx \num{0.80}\,\pi$,
        two fixed points vanish in a saddle-node bifurcation.
        The three markers correspond to
        the trajectories shown in Fig.~\ref{fig:potential}.
        (b)~Rate constant \kFloq\ as a function of driving amplitude $\hat{x}$
        for three different driving frequencies $\omega_x$.
        Only a single fixed point (\ie, periodic trajectory)
        was found per set of parameters considered here.
        Orange circles indicate
        the parameter sets used in Fig.~\ref{fig:trajectories}
        while black diamonds mark each curve's maximum.
        The vertical gray line acts as a guide to the eye.}
    \label{fig:param_dep_rates}
\end{figure}

These two examples demonstrate that
the dynamics of trajectories on or near the \ac{NHIM}
can be drastically altered through modification of the driving parameters.
This is summarized in Fig.~\ref{fig:param_dep_rates}
through the calculation of the decay rates \kFloq\
as a function of the driving frequency and amplitude.
As these are only one-dimensional sections
through the two-dimensional space of possible driving parameters,
they serve here as examples only.
Specifically, they are not meant to represent an exhaustive or exclusive set.
As such, any extremal values in one of these sections
may not necessarily be extremal in the full parameter space.
Nevertheless, the existence of such extrema
in sections of parameter space
is enough to demonstrate that these systems are sensitive to the driving.

The bifurcation observed in Fig.~\ref{fig:energy_rate_psos}
is visible in Fig.~\ref{fig:param_dep_rates}(a).
At larger driving frequencies, there exists a single elliptic fixed point.
When lowering $\omega_x$, its rate constant \kFloq\ steadily increases.
Around $\omega_x = \num{0.80}\,\pi$,
two new fixed points with lower values of \kFloq\
emerge in a saddle-node bifurcation.
Furthermore, a comparison with the trajectories from Fig.~\ref{fig:potential}
suggests that high rates are accompanied
by large motion in the orthogonal mode $(y, v_y)$.

We demonstrated through Fig.~\ref{fig:trajectories} that there exists
a minimum in stability for medium driving.
This manifests itself in Fig.~\ref{fig:param_dep_rates}(b)
by means of a maximum in \kFloq.
When varying $\omega_x$, this extremum persists qualitatively the same,
differing mainly in position and height.
The latter can be connected to the slope in Fig.~\ref{fig:param_dep_rates}(a).
Note that all curves, independent of $\omega_x$,
must meet at $\kFloq(\hat{x} = 0) \approx \num{2.762}$
since a vanishing amplitude is equivalent to the static case.


\subsection{Reaction geometry}
\label{sec:results/geometry}

\begin{figure}[t]
    \includegraphics[width=\columnwidth]{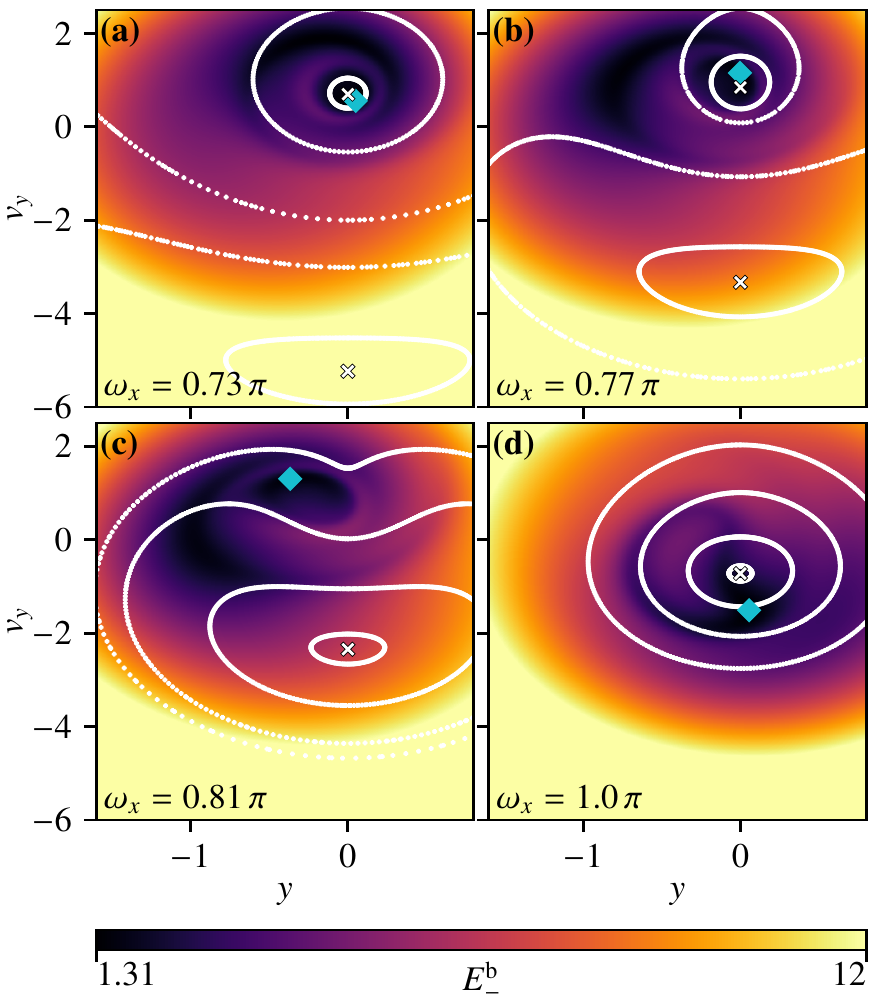}
    \caption{%
        Local threshold energy $\ESrc = \lim_{t \to -\infty} E_\traj$
        of trajectories \traj\
        started from position $(y, v_y)$ close to the \acs{NHIM}
        at crossing time $t_0 = 0$
        and propagated back to the initial time $t \to -\infty$.
        A \ac{PSOS} showing the structure of the \ac{NHIM}
        is overlaid for comparison.
        Small white crosses mark the positions of elliptic fixed points.
        The diamond marker in each panel is placed at the global spacial minimum
        $\ESpMin(t_0) = \min_{y, v_y} \ESrc(y, v_y, t_0)$.
        The panels (a) through (d) were calculated
        for different driving frequencies $\omega_x$
        as indicated in the bottom left of each panel.
        The driving amplitude is $\hat{x} = \num{0.4}$ for all panels.
        The color scale is capped at $\ESrc = \num{12}$.}
    \label{fig:startenergy}
\end{figure}

The results reported in the previous sections
relate only to the dynamics on the \ac{NHIM}.
Making predictions about real chemical reactions, however,
requires us to connect to the dynamics off the \ac{NHIM}.
More specifically, we need to address
when and how the \ac{NHIM} can influence reactive trajectories,
\ie, those connecting the reactant to the product basins.

The \ac{NHIM} represents---by construction---%
the minimum energy a trajectory needs for any given set of orthogonal modes $(y, v_y)$ to cross the \ac{DS}.
It is therefore natural to assume that
a significant portion of reactants in a thermally distributed ensemble
would pass close to the \ac{NHIM} while reacting.
Additionally, for reasons of continuity,
we can expect these trajectories to behave similarly to those on the \ac{NHIM}
for some finite time.
This provides a possible connection
between the dynamics on and off the \ac{NHIM}.

In a driven system,
reactants may gain or lose energy while climbing the potential barrier.
A trajectory's energy $E_\traj$ very close to the \ac{NHIM}
can thus differ from its initial energy
$\ESrc = \lim_{t \to -\infty} E_\traj$
in the reactant basin.
The structure of the \ac{NHIM} can be connected to
the reactant basins through propagation back in time.
For each fixed orthogonal mode $(y, v_y)$ and time $t_0$,
we first obtain the position $(x, v_x)$ of the \ac{NHIM}.
A shift of this point by $\Delta v_x = +\num{e-5}$
yields a point on a reactive trajectory which closely passes the \ac{NHIM}.
We then propagate the trajectory backward in time
until we are sufficiently far away from the moving barrier.
The trajectory's energy \ESrc\ at this early time%
---which we refer to as the \emph{local threshold energy}---%
will then be approximately conserved.
Through sampling $(y, v_y)$, we then obtain
the distribution of local threshold energies and
the corresponding initial points in phase space of the reactive trajectories.
Figure~\ref{fig:startenergy}
reports the results of this calculation for crossing time $t_0 = 0$
at four driving frequencies $\omega_x$
around the bifurcation shown in Fig.~\ref{fig:param_dep_rates}.
For comparison,
the structure of the \ac{NHIM} as revealed by a \ac{PSOS}
has been overlaid in each case.

The \ac{PSOS} reveals two elliptic fixed points on the \ac{NHIM}
at the driving frequencies below the bifurcation
[\cf\ Figs.~\ref{fig:startenergy}(a) and \ref{fig:startenergy}(b)].
Although the lower one cannot be seen
in the structure of the local threshold energy,
there is a correlation between \ESrc\ and the upper fixed point.
This is consistent with the fact that
the lower fixed point is associated with higher decay rates
as shown in Fig.~\ref{fig:param_dep_rates}.
Trajectories consequently spend less time near the \ac{NHIM},
and we expect less correlation with the dynamics on the \ac{NHIM}.
Conversely,
there is a very good match between the \emph{global threshold energy}
\begin{equation}
    \ESpMin(t_0) = \min_{y, v_y} \ESrc(y, v_y, t_0)
\end{equation}
and the position of the upper fixed point, \ie,
the trajectory on the \ac{NHIM} with the least average energy
(\cf\ Fig.~\ref{fig:energy_rate_psos}).

Closer to the bifurcation,
we find the structure of \ESrc\ starting to change
[\cf\ Figs.~\ref{fig:startenergy}(b) and \ref{fig:startenergy}(c)].
Low-energy regions seem to flow out
in a counter-clockwise spiral-like structure.
The minimum \ESpMin, however, stays near the upper fixed point.
It only starts to move once this fixed point
disappears in the bifurcation.
Following a counter-clockwise trajectory itself,
it moves down towards the remaining fixed point,
slowly converging for increasing driving frequency $\omega_x$
[\cf\ Fig.~\ref{fig:startenergy}(d)].

If we assume initial energies of a reactant ensemble to be thermally distributed,
then we can expect most of these reactants
to react via paths related to low-\ESrc\ regions at the crossing time $t_0$.
The bifurcation, therefore, should change
the geometry of the reaction dynamics at least qualitatively.
This change, as indicated by the
movement of the global spacial minimum of the threshold energy
$E_{\rm th}^{\rm b}(t_0)$, appears to be smooth across the
bifurcation.
As a consequence, we can anticipate that the
reaction rates to be presented in the next section will
not exhibit a discontinuity around the bifurcation.


\subsection{Reaction probability}
\label{sec:results/probability}

\begin{figure}[!t]
    \includegraphics[width=\columnwidth]{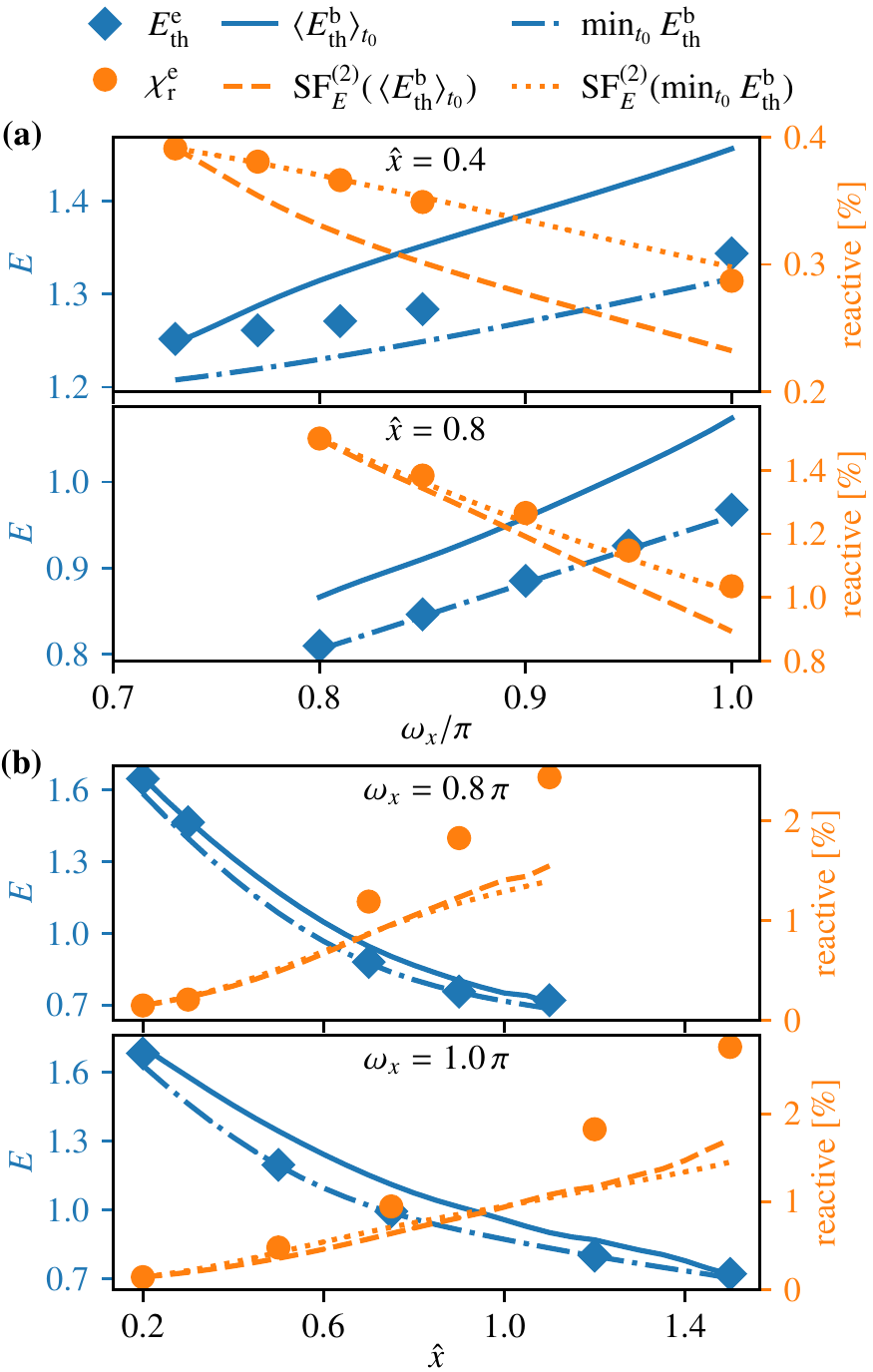}
    \caption{%
        Ensemble threshold energy \EEnsMin\ (diamond markers)
        and fraction of reactive trajectories \reacEns\ (circle markers)
        for the ensemble calculation
        described in Sec.~\ref{sec:results/probability}.
        Each of the four panels varies either
        (a)~driving frequency $\omega_x$ or (b)~driving amplitude $\hat{x}$
        while the other parameter is kept fixed.
        For comparison, calculations based on the global minimum
        $\ESpMin(t_0) = \min_{y, v_y} \ESrc(y, v_y, t_0)$
        of the local threshold energy \ESrc\ akin to Fig.~\ref{fig:startenergy}
        are shown.
        Specifically, the average \ESpMinTmAvrg\
        and minimum \ESpMinTmMin\
        in crossing time $t_0$ are shown
        with solid and dash-dotted lines, respectively.
        Evaluating the survival function \survival{E}{1}
        of a one-dimensional Maxwell--Boltzmann distribution in energy space
        and scaling the results to fit the first ensemble data point
        yields the dashed and dotted lines, respectively.
        Parameter ranges differ because of numerical stability.}
    \label{fig:reac_prob}
\end{figure}

We now address the degree to which reaction rates%
---not just decay rates---%
can be obtained from the structure of the \ac{NHIM}.
Following Farkas and Kramers~\cite{farkas27, kram40, rmp90, rsh99},
the reaction rate is determined by the ratio of the
reactive flux across a
\ac{DS} divided by the reactant population
at steady-state conditions.
This presumes a boundary condition in which the reactants are continuously populated
at the well according to an equilibrium condition.
Here we assume that the reactants are initially thermally distributed,
and set the initial distribution in velocities to be that of Boltzmann
at temperature, $T$, and located in the reactant basin far from the
\ac{NHIM}.
The system is then propagated semi-microcanonically---%
viz., including external driving but neglecting friction and noise.
This corresponds to a system which is very weakly coupled to an external bath.
The rates that one would obtain in this way
are therefore good approximations in cases in which
the rate is fast compared to the dissipation.

For numerical expedience, here we obtain the reaction fraction rather
than the rates using the flux over population approach.
The reactant fraction
is the fraction of particles that react---before their
first return to the reactant basin---to products given the initial
distribution.
The reference ensemble simulation is constructed as follows.
For every set of parameters,
we initialize an ensemble of \num{e7} reactants at position $(x, y) = (-8, -1)$,
that is, far from the saddle on the minimum energy path.
Velocities $v_x$ and $v_y$ are chosen according to
a Maxwell--Boltzmann distribution
of temperature $\kB T = \num{0.4} \ll E_\mathrm{b}$.
Negative velocities result
in trajectories that cannot react
because the reactant basin is unbounded.
We thus include only positive velocities
$v_x \ge 0$ by taking the absolute value.
The initial time is chosen based on a uniform random distribution.
Each reactant is then propagated forward in time
until it leaves the reaction region.
Trajectories passing $x < -8$ are classified as nonreactive
and those passing $x > +4$ as reactive.
Besides the fraction of reactive trajectories \reacEns,
we additionally record
the minimal initial energy \EEnsMin\ of the reactive subensemble,
referred to as the \emph{ensemble threshold energy}.
The results for various values of
the driving parameters $\omega_x$ and $\hat{x}$
are shown as circle and diamond markers in Fig.~\ref{fig:reac_prob}.

Alternatively, we can consider the dynamics on the \ac{NHIM} directly
using the spacial minimum \ESpMin\
as the effective minimum barrier height;
see Sec.~\ref{sec:results/geometry}.
We employ modern global minimization routines
to make the determination of \ESpMin\ as efficient as possible.
Specifically, we use simplicial homology global optimization~\cite{Endres2018}
with Sobol' sampling~\cite{Sobol1967}
and the Nelder--Mead simplex method~\cite{NelderMead1965} for local optimization
as implemented in the Python library SciPy~\cite{SciPy2020}.
The resulting \ESpMin\ is still dependent on the crossing time $t_0$.
To account for this fact,
we consider both the average \ESpMinTmAvrg*\
and the minimum \ESpMinTmMin\ in $t_0$ going forward.
Both quantities are shown in the left axes of Fig.~\ref{fig:reac_prob}
for multiple driving parameter ranges.
Unsurprisingly, the minimum ensemble energy \EEnsMin\
is close to but always larger than \ESpMinTmMin.

The most straightforward way
to obtain a reaction probability from a barrier height
is by evaluating the ensemble's complementary cumulative distribution function%
---also known as the survival function.
In energy space, Maxwell--Boltzmann ensembles follow a $\chi^2$ distribution
with argument $2 E / (\kB T)$.
Here, we report the survival probability \survival{E}{2}
according to the energy distribution
over the two-dimensional configuration space, $x$ and $y$.
Curiously, the agreement in the reactive probability (not shown here)
was better in the cases reported in Fig.~\ref{fig:reac_prob}(b)
when we evaluated the survival probability
using only the distribution over the reactive degree of freedom, $x$.
For reactive trajectories, this circumstance suggests that
the nonlinear coupling between $x$ and $y$ is not strong enough
to lead to a significant energy exchange
between the reaction coordinate and the orthogonal mode.
The dynamics on the \ac{NHIM}, however,
is definitely affected by the nonlinear coupling
as shown by the bifurcation in Fig.~\ref{fig:param_dep_rates}.
Finally, we calibrate the resulting curve
by linearly scaling it to match the first value of the ensemble calculation.
The result is shown in the right axes of Fig.~\ref{fig:reac_prob}.
There is a clear correlation between
\reacEns\ and the survival function of \ESpMinTmMin\
with very good agreement for $\hat{x} \lesssim \num{0.8}$.
The average \ESpMinTmAvrg*, on the other hand,
yields worse results in most cases.
It can only slightly beat \ESpMinTmMin\
for very high driving amplitudes $\hat{x}$.
That is, it appears that the deviations in the
reactive percentage between the use of
the global reactive flux and the \ac{NHIM}-based approaches
arises because the globality presumed in the latter
begins to break down as the particles are driven
harder and farther away from the reactive region.


\section{Concluding remarks}
\label{sec:conclusion}

In this paper, we have demonstrated that
decay rates and the reaction geometry can be manipulated by external driving.
Based on this, we have found a connection between
properties of the \ac{NHIM} and properties of reacting trajectories.
This, in turn, has allowed us to predict reaction probabilities
without having to propagate large ensembles for each set of parameters,
providing further insights into the dynamics of chemical reactions.
In the future, these results could be used
to control and optimize the reaction rate of chemical reactions.
To achieve this goal, however,
it is required to extend the methods discussed here
to models explicitly describing particular chemical reactions.
Promising candidates include the isomerization reactions of
LiCN~\cite{hern08g, hern12e, hern14j, hern16c},
KCN~\cite{borondo13a, borondo18a},
and ketene~\cite{gezelter1995, hern13c, wiggins14b, hern14e, hern16d}.
Additionally, the results have to be extended
to include noise and friction (\ie\ Langevin dynamics)
in order to be applicable to real chemical reactions.


\section*{CRediT authorship contribution statement}

\textbf{Johannes Reiff:}
    Methodology,
    Software,
    Validation,
    Formal analysis,
    Investigation,
    Data Curation,
    Writing -- Original Draft,
    Writing -- Review \& Editing,
    Visualization.
\textbf{Robin Bardakcioglu:}
    Methodology,
    Software,
    Investigation.
\textbf{Matthias Feldmaier:}
    Methodology,
    Writing - Original Draft,
    Visualization.
\textbf{Jörg Main:}
    Conceptualization,
    Methodology,
    Resources,
    Writing -- Original Draft,
    Writing -- Review \& Editing,
    Supervision,
    Project administration,
    Funding acquisition.
\textbf{Rigoberto Hernandez:}
    Conceptualization,
    Writing -- Review \& Editing,
    Project administration,
    Funding acquisition.


\section*{Declaration of competing interest}

The authors declare that they have
no known competing financial interests or personal relationships
that could have appeared to influence the work reported in this paper.


\section*{Acknowledgments}

The German portion of this collaborative work was partially supported
by the Deutsche Forschungsgemeinschaft (DFG) through Grant No.~MA1639/14-1.
The US portion was partially supported
by the National Science Foundation (NSF) through Grant No.~CHE~1700749.
MF is grateful for support
from the Landes\-graduierten\-förderung of the Land Baden-Württemberg.
This collaboration has also benefited from support
by the European Union's Horizon 2020 Research and Innovation Program
under the Marie Skłodowska-Curie Grant Agreement No.~734557.


\bibliographystyle{elsarticle-num-names}
\bibliography{paper-q25letter}
\end{document}